\documentclass[conference,a4paper]{IEEEtran}
\usepackage[utf8]{inputenc} 
\usepackage[T1]{fontenc}
\usepackage{url}              % provides \url{...}
\usepackage{cite}             % improves presentation of citations

\usepackage[cmex10]{amsmath}  % Use the [cmex10] option to ensure complicance
\usepackage{mleftright}       % fix to wrong spacing of \left-,
\mleftright                   % \middle- \right-commands 

\usepackage{graphicx}         % provides \includegraphics{...} to
\usepackage{booktabs}         % fixes poor spacing in tables and
\usepackage{amsmath,amssymb,amsfonts,dsfont,subfigure}
\usepackage{cite, color,algorithm, algpseudocode, comment}
\usepackage{braket}

\newtheorem{theorem}{Theorem}
\newtheorem{definition}{Definition}
\newtheorem{lemma}{Lemma}

\newtheorem{remark}{Remark}

\newtheorem{example}{Example}

\newtheorem{claim}{Claim}

\newcommand{\blank}[1]{}

\newcommand{\A}{\mathbf{A}} 
\newcommand{\B}{\mathbf{B}}
\newcommand{\NNL}{\nonumber\\}
\newcommand{\NN}{\nonumber}
\newcommand{\bfH}{\mathbf{H}}
\newcommand{\bfh}{\mathbf{h}}
\newcommand{\calI}{\mathcal{I}}

\newcommand\blfootnote[1]{%
  \begingroup
  \renewcommand\thefootnote{}\footnote{#1}%
  \addtocounter{footnote}{-1}%
  \endgroup
}

\begin{document}

%\title{ISIT 2023 Paper Template\\Please Capitalize Important Words in Title} 
\title{Coded matrix computation with gradient coding}
%\title{Recovery threshold does not imply the computation speed} 

%%%%%%
\author{%
  \IEEEauthorblockN{Kyungrak Son} 
  \IEEEauthorblockA{Institute of New Media and Communications,\\
                    Seoul National University,\\
                    Seoul 08826, South Korea,\\
                    kyungrakson@snu.ac.kr}
  \and
  \IEEEauthorblockN{Aditya Ramamoorthy} 
  \IEEEauthorblockA{Dept. of Electrical and Computer Eng.,\\
                    Iowa State University,\\
                    Ames, IA 50011, U.S.A.,\\
                    adityar@iastate.edu}                    
}

\maketitle
\blfootnote{The work of K. Son was supported in part by Basic Science Research Program through the National Research Foundation of Korea(NRF) funded by the Ministry of Education(grant NRF-2021R1A6A3A01086690). The work of A. Ramamoorthy was supported in part by the National Science Foundation (NSF) under Grant CCF-1910840 and Grant CCF-2115200.}

%%%%%
%% Abstract: 
%% If your paper is eligible for the student paper award, please add
%% the comment "THIS PAPER IS ELIGIBLE FOR THE STUDENT PAPER
%% AWARD." as a first line in the abstract. 
%% For the final version of the accepted paper, please do not forget
%% to remove this comment!
%%
\begin{abstract}
    Polynomial based approaches, such as the Mat-Dot and entangled polynomial codes (EPC) have been used extensively within coded matrix computations to obtain schemes with good recovery thresholds. However, these schemes are well-recognized to suffer from poor numerical stability in decoding. Moreover, the encoding process in these schemes involves linearly combining a large number of input submatrices, i.e., the encoding weight is high. For the practically relevant case of sparse input matrices, this can have the undesirable effect of significantly increasing the worker node computation time.
    In this work, we propose a generalization of the EPC scheme by combining the idea of gradient coding along with the basic EPC encoding. Our technique allows us to reduce the weight of the encoding and arrive at schemes that exhibit much better numerical stability; this is achieved at the expense of a worse threshold. By appropriately setting parameters in our scheme, we recover several well-known schemes in the literature. Simulation results show that our scheme provides excellent numerical stability and fast computation speed (for sparse input matrices) as compared to EPC and Mat-Dot codes.
    
    %provides a general model which can control the number of summation terms (i.e., weights) of the encoded submatrices. Thereby, we are able to control the numerical stability as we want. From the simulation, the proposed scheme provides excellent numerical stability and fast computation speed with the tradeoff of recovery threshold compared to the conventional scheme.  
\end{abstract}

% \begin{abstract}
% It is well-recognized that the MatDot code and entangled polynomial code suffer from poor numerical stability even for moderate value of matrix partitioning size. 
% %
% To resolve this issue, in this paper, we propose a distributed matrix multiplication scheme, which utilizes the gradient code on top of the entangled polynomial code. Our scheme provides a general model which can control the number of summation terms (i.e., weights) of the encoded submatrices. Thereby, we are able to control the numerical stability as we want. From the simulation, the proposed scheme provides excellent numerical stability and fast computation speed with the tradeoff of recovery threshold compared to the conventional scheme.  
% \end{abstract}

\section{Introduction}
Large scale matrix computations are at the heart of various machine learning and optimization problems. In many of these problems, the size of the underlying matrices requires the usage of distributed computing, where the overall job is divided into smaller tasks that can be executed in parallel over multiple workers. However, straightforward task assignments can result in situations where the job execution time is limited by the speed of the slowest worker. This is especially problematic in cloud computing scenarios where workers are well-recognized to exhibit appreciable variance in computing speeds \cite{das2019random}.
 
{\bf Background:} The field of coded matrix computation \cite{lee2018speeding,yu2017polynomial, yu2020straggler,duttaCG16,ramamoorthyDTMag20} aims at leveraging ideas from coding theory to improve the overall job execution time within distributed clusters. Given matrices $\A \in \mathbb{R}^{\beta \times \alpha}$ and $\B \in \mathbb{R}^{\beta \times \gamma}$, suppose that we are interested in computing $\A^T \B$. In coded computation, a designated central node performs a block decomposition of $\A$ and $\B$ and assigns encoded submatrices of them to the worker nodes. The task of the worker nodes is now to compute the product of these encoded matrices.
For carefully designed schemes, it can be shown that the desired result can be decoded as long as {\it any} $\tau$ worker nodes return their results. Thus, the job execution time is not dominated by slow workers. $\tau$ is known as the threshold of the scheme.

More recently, it has been recognized \cite{8849451,8849395,das2020coded, ramamoorthy2019numerically, 8919859,kiani2018exploitation,Son2021,Son2022} that there are other metrics that are also of interest within coded matrix computation. The work of \cite{c3les,8849451,ramamoorthy2019numerically,das2020coded,8849468} has demonstrated that several of the original polynomial-based schemes suffer from the problem of numerical instability (i.e., computation error caused by distributing the computation). In particular, the decoded result in these schemes can be essentially useless even for clusters with thirty nodes or more. Furthermore, in several settings, the input matrices $\A$ and $\B$ are sparse. Note that the encoding process typically combines a number of different submatrices of $\A$ (and $\B$); we refer to this as the encoding weight of the scheme. This encoding can significantly increase the number of non-zero entries in the encoded matrices. This in turn will have the undesired effect of increasing the worker node computation time \cite{das2020coded,dasunifiedtreatment,wang2018coded}. Thus, coded computation schemes that have small encoding weights are of interest. Other metrics include how well a given scheme leverages partial computations performed by the worker nodes \cite{kiani2018exploitation,das2020coded,dasunifiedtreatment,c3les}.

Within coded computation, the central node first performs a block-decomposition of $\A^T$ and $\B$ as follows.

\begin{align}
\A^T
%\!\!=\!\!
&=
\left[
\begin{matrix}
	\A_{0,0}^T & \!\!\!\cdots \!\!\!& \A_{p-1,0}^T  \\
	\vdots & \!\!\!\ddots\!\!\! & \vdots  \\
	\A_{0,m-1}^T &\!\!\!\cdots \!\!\!& \A_{p-1,m-1}^T 
\end{matrix}
\right], \text{and} \nonumber\\
\B
%\!=\!\!
&=
\left[
\begin{matrix}
	\B_{0,0}  & \!\!\!\cdots\!\!\!& \B_{0,n-1} \\
         \vdots & \!\!\!\ddots\!\!\! & \vdots  \\
	\B_{p-1,0}  &\!\!\!\cdots\!\!\! & \B_{p-1,n-1}
\end{matrix}
\right]. \label{eq:block_decomp}
\end{align}
Each worker node is allowed to store the equivalent of $1/pm$-fraction of $\A$ and $1/pn$-fraction of $\B$.
The overall idea is to encode the submatrices of $\A$ and $\B$ and assign the worker nodes the task of computing the product of these encoded submatrices, such that the central node can decode if enough tasks are completed.

{\bf Related Work:}
In polynomial-based schemes \cite{dutta2019optimal,yu2020straggler, salman_secured}, the encoding functions are polynomial evaluation maps. Upon multiplication of the encoded matrices, the desired terms appear as coefficients of certain monomials and the other coefficients are treated as interference. If enough worker nodes return their results, there are enough evaluation points so that the polynomial can be interpolated and the desired terms and hence $\A^T \B$ can be recovered. 

In particular, the Mat-Dot code \cite{dutta2019optimal} applies in the setting when $m=n=1$ and arbitrary $p$ and has recovery threshold of $2p-1$. The entangled polynomial code (EPC) \cite{yu2020straggler} applies for any $m,n$ and $p$ and has threshold of $pmn+p-1$. The decoding process in both cases requires interpolating polynomials of degree $2p-2$ and $pmn+p-2$ respectively. There have been several works that have examined the case of $p=1$. 

The issue of numerical stability has been examined in several works. For instance, \cite{8849468} works within a different basis set of polynomials. In \cite{ramamoorthy2019numerically}, the authors presented a technique that exploits the properties of rotation and circulant permutation matrices for improved numerical stability and in \cite{8919859,das2020coded}, the authors used random linear combinations for the encoding. Low weight encodings were considered in \cite{das2020coded,dasunifiedtreatment} that also demonstrated a scheme that continues to have the optimal threshold. Finally, techniques that leverage partial stragglers have also been investigated in several works \cite{kiani2018exploitation,das2020coded,dasunifiedtreatment,c3les}. We note here that for large values of $p,m$ and $n$, the numerical instability issue with the Mat-Dot and EP code approaches is especially acute. In addition, as we will see their encoding weights are also high, rendering them unsuitable for sparse input matrices. 

{\bf Main Contributions:}  
\begin{itemize}
    \item In this work, we present a coded computation scheme that allows us to trade-off the interpolation degree of the reconstructed polynomial \& encoding weight with the recovery threshold for EP and Mat-Dot codes. By operating on this tradeoff we can arrive at schemes that are significantly more stable numerically and suitable for sparse input matrices. Our schemes proceed by combining the idea of gradient coding (GC) \cite{tandon2017gradient} and the structure of the EP codes. We calculate the recovery threshold of our scheme.
    \item We show that \cite{yu2020straggler} and \cite{Charalambides2021} can be viewed as two extremes of the proposed scheme depending on the choice of parameters. Thus, our proposed scheme is a generalization of these schemes. 
    \item Extensive simulation results corroborate our theoretical findings. %From the simulation, 
\end{itemize}
We point out that there is a related work that utilizes GC for coded matrix computations in \cite{Charalambides2021}. However, we note that this paper is different from our paper since \cite{Charalambides2021} focuses more on designing numerically stable GC using binary coefficients and does not analyze the recovery threshold. We discuss this in more detail in Sections \ref{sub:discussion} and \ref{sec:sim_results}.

{\bf Notation:} For integers $a, b$, the notation $a | b$ denotes that $a$ divides $b$. For a set of vectors $\mathcal{V}$, $\mathrm{span}(\mathcal{V})$ denotes the span of the vectors (i.e., set of all linear combinations of the vectors) in $\mathcal{V}$. If $\mathbf{A}$ is a set of integers then $\mathbf{A} \mod \ell$ denotes $\mathbf{A}$ with all elements reduced modulo $\ell$.

%$\mathrm{Card}(S)$ is the cardinality of the set $S$. For the set $S_1,S_2$, $S_1\backslash S_2 = \{x|x \in S_1, x \notin S_2\}$.

%
%\aditya{Is this needed now?} \blue{We generally assume that all of the index terms in vector (or matrices) are under the modulo operation (i.e., index $a+b$ equals to $a+b~ (\textrm{mod~} c)$ for the length $c$ vector.)}

\section{sparsity controlled distributed matrix multiplication with general matrix partitions}
% Before starting, we briefly introduce the encoding and decoding matrix of gradient coding, which will be utilized in the paper. 
% \begin{definition} \label{def:grad}{(Conditions of gradient coding \cite{Tandon2017,Charalambides2021})} For a given number of nodes $\eta$ and target number of stragglers $\kappa$, the following conditions should be satisfied in order to construct a gradient code robust to any $\kappa$ full stragglers. %[\mathbf{g}_1^T \cdots \mathbf{g}_{\binom{N}{N-s}}^T]^T 
% \begin{itemize}
%     \item Each row of the encoding matrix $\mathbf{H}=[\mathbf{h}_1, \cdots, \mathbf{h}_\eta] \in \mathbb{R}^{\eta \times \eta}$ should be chosen from the $\eta-\kappa$ dimensional subspace containing all ones vector. i.e, for every $I \in \{1,\cdots, \eta\}$ with the size $\eta-\kappa$, 
%     \begin{align}
%         \mathds{1}_{1\times \eta} \in \mathrm{span}(\{\mathbf{h}_i|i \in I\}).
%         \label{eqn:span}
%     \end{align}
%     The precise construction algorithm of $\mathbf{H}$ is given in \cite{Tandon2017}.
%     \item Thus, for an arbitrary subset $S \in \{1,\cdots, \eta\}$ with the size $\kappa$, there exist a decoding vector $\mathbf{g}^T$ having zeros in $S$ and satisfies 
%     \begin{align}   \mathbf{g}^T\mathbf{H}=\mathds{1}_{1 \times \eta}
%         \label{eqn:all_ones}
%     \end{align}
% \end{itemize}
% \end{definition}
% Now, we start with a motivating example.

\begin{figure*}[t!]
		\centering	
  \subfigure[A simple scheme with two independent MatDot codes.]{\includegraphics[width=.98\columnwidth]{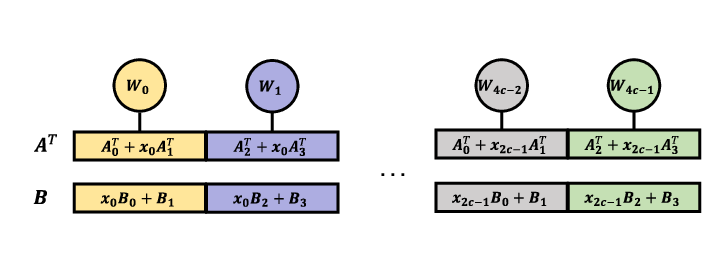} }
  \subfigure[A group $\mathcal{G}_i$ of the proposed scheme.]{\includegraphics[width=.98\columnwidth]{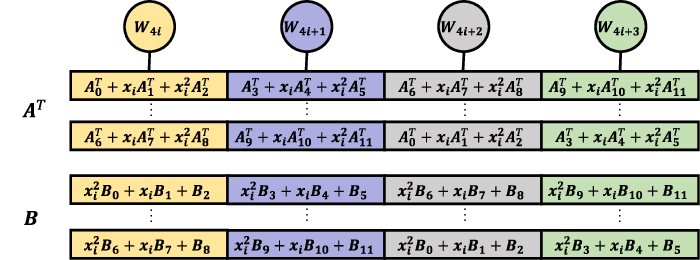} }
	\vspace{0.1in}
	\caption{Task assignment in Example 1. There are $N=4c$ workers for distributed matrix computation with storage size $\gamma_A=\gamma_B=$ $1/4$. In Fig. 1(b), each worker has three encoded $\A$ and $\B$ assignments. Owing to space limitations, we use vertical dots to denote the missing encoded assignments.}
	\label{fig:grad_assignment}
\end{figure*}
\begin{definition} \label{def:grad} {\it Gradient Coding matrix.}
Let $\bfH$ be a $\eta \times \eta$ matrix, with its rows denoted $\bfh_i, i = 0, \dots, \eta-1$. We say that $\bfH$ is a gradient coding matrix with parameters $\eta$ and $\kappa$ if it has the following properties.
\begin{itemize}
    \item[(i)] It has cyclically shifted rows and each row has $\kappa + 1$ non-zero entries. Let $\calI_i = \{i, i+1, \dots, i+\kappa\} \mod \eta$. Row $\bfh_i = [h_{i,0}~ h_{i,1}~ \dots~h_{i,\eta-1}]$ is such that $h_{i,j} \neq 0$, if and only if $j \in \calI_i$.
    \item[(ii)] The all-ones row vector is contained in the span of any $\eta - \kappa$ rows of $\bfH$, i.e., for $J \subset \{0, \dots, \eta-1\} \text{~with~} |J| = \eta - \kappa$ we have,
        \begin{align}
        \mathds{1}_{1\times \eta} \in \mathrm{span}(\{\mathbf{h}_i|i \in J\}).
        \label{eqn:span}
    \end{align}
\end{itemize}

\end{definition}

\subsection{Motivating example}

\begin{example}
    	Suppose there are $N=4c ~(c\geq 5)$ workers. Henceforth, let $\gamma_A$ and $\gamma_B$ denote the storage fraction of matrices $\A$ and $\B$. We assume that each worker can store the equivalent of $\gamma_A=\gamma_B=1/4$ fractions of matrices $\A \in \mathbb{C}^{\beta \times \alpha}$ and $\B \in \mathbb{C}^{\beta\times \gamma}$, respectively. The MatDot code \cite{dutta2019optimal} where $\A^T$ and $\B$ are decomposed into four block-columns is applicable here $(m=n=1, p = 4)$ and is resilient to $N-7$ stragglers. In this approach the encoded $\A$ and $\B$ submatrices involve linear combinations of all the respective submatrices, and decoding requires interpolating a polynomial of degree $6$. 

        %\aditya{provide motivation for low-weight encoding and for low-degree interpolation.}
        Now suppose that we are interested in a scheme where weight of the encoding matrices (both $\A$ and $\B$) is two. In this case, a simple technique is to work with two independent MatDot schemes each with $(m=n=1, p' = 2)$. We first partition $\A^T$ and $\B$ as $\A^T=[\A_0^T \cdots \A_3^T]$ and $\B=[\B_0 \cdots \B_3]^T$. Then,
        we divide the workers into $2c$ groups where each group consists of two workers such that the $w=0,1$-th worker of the group stores 
        \begin{align*}
        \bar{\A}^T(w,x_i) &= 
  \sum_{l=0}^{1} x_i^l \A_{2w+l}^T  \textrm{~~and~~} \bar{\B}(w,x_i) = 
  \sum_{l=0}^{1} x_i^{1-l} \B_{2w+l}.
        \end{align*}
    as illustrated in Fig. \ref{fig:grad_assignment}(a). The value of $x_i$ is fixed for a group. It is not hard to see that the recovery threshold of above scheme is $2c+3$. Since the product of $\bar{\A}^T(w,x_i)$ and $\bar{\B}(w,x_i)$ yields a degree-2 polynomial, we can decode as long as we obtain three evaluations of each of the two relevant polynomials corresponding to $w=0,1$. Thus, we cannot decode when we have all the results, e.g., of the  polynomial for $w=0$ from all of $2c$ groups and the result of the polynomial for $w=1$ from at most two groups. Thus, the recovery threshold becomes $2c+3$.  

    The situation differs when we are interested in schemes where the encoding weight, e.g., is three. In this case, the encoding weight does not divide $p=4$. Thus, a simple scheme as the one discussed above cannot be found in a straightforward manner. Instead, consider the following scheme. We partition $\A^T$ and $\B$ into  $12$ submatrices denoted
 
 %This is well-recognized to be numerically unstable for even moderate values of $p$. \aditya{Talk about low-weight combinations. May move this discussion to the Introduction} 
 
 %In this work we propose to use gradient coding along with polynomial encoding for coded matrix computation. As we will see, this will reduce the degree of the underlying polynomial that needs to be interpolated. Furthermore, our scheme allows us to work with lower-weight encoding than those used in MatDot and EP codes. \aditya{will reword}

 %the condition number of MatDot code becomes $O(q^{q-7+5.5})$ with a prime $q>7$ using \cite{Ramamoorthy2022}, which is indeed a large number, which causes a great loss in the numerical stability. 
 %To solve this numerical stability issue, we consider a scheme using gradient code \cite{Tandon2017} on top of the fractions of MatDot code and show that this simple fractioning is indeed powerful in terms of numerical stability. 

%Consider the following scheme: the central node first partitions the matrices $\A^T$ and $\B$ into submatrices 
	\begin{align*}
		\A^T
		=
		\left[
		\begin{matrix}
			\A_{0}^T & \cdots & \A_{11}^T
		\end{matrix}
		\right],~\text{and~}
		\B
		=
		\left[
		\begin{matrix}
			\B_{0} \\
			\vdots \\
			\B_{11} 
		\end{matrix}
		\right].
	\end{align*}
	Since the partitioned submatrices have $1/12$-th the size of matrices $\A$ and $\B$, we can store three of them in each worker while still respecting the storage constraint. Now, consider a group of four workers $\mathcal{G}_i = \{4i, \cdots, 4i+3\}$ for $i=0, \dots, c-1$, where the worker $4i+w$ stores 
     \begin{equation}
    %\label{eqn:example1_encoded_mat}
     \begin{aligned} 
\bar{\A}^T(x_i,\tilde{p}) &= 
  \sum_{l=0}^{2} x_i^l \A_{3\tilde{p}+l}^T
%h_{ww}\A_{w\tilde{m}}^T+h_{w(w+1)}x_i\A_{(w+1)\tilde{m}}^T,
	\end{aligned}
 \end{equation}	for all $\tilde{p}\in \mathcal{P}_\mathsf{p}=\{w, \dots, w+2\} \mod 4$ (note that the index $\tilde{p}$ depends upon $w$) and 
     \begin{equation}
    \label{eqn:example1_encoded_mat}
     \begin{aligned} 
		\bar{\B}(x_i,\tilde{p}) &= 
  \sum_{l=0}^{2} x_i^{2-l} \B_{3\tilde{p}+l}
%x_i\B_{w\tilde{n}}+\B_{(w+1)\tilde{n}},
	\end{aligned}
 \end{equation}	
for all $\tilde{p}\in \mathcal{P}_\mathsf{p}$. Here, $x_i$ is the same for all workers in the group.
Next, we choose a matrix $\mathbf{H}$ that satisfies \eqref{eqn:span} with the parameters $\eta=4$ and $\kappa=2$ and assign each worker the task of computing
    \begin{align}	\mathbf{C}_w(x_i) &\!=\! \sum_{\tilde{p}=w}^{w+2}h_{w,\tilde{p}}\bar{\A}^T(x_i,\tilde{p})\bar{\B}(x_i,\tilde{p}) \label{eq:eg_w_index_mod}\\
    &\!\stackrel{(a)}{=}\! \sum_{\tilde{p}'=0}^{3}h_{w,\tilde{p}'}\bar{\A}^T(x_i,\tilde{p}')\bar{\B}(x_i,\tilde{p}'), \nonumber
    \end{align}
where $h_{i,j}$ is the $i$-th row and $j$-th column of the matrix $\mathbf{H}$. The summation indices in \eqref{eq:eg_w_index_mod} are reduced modulo-4.

  %   \begin{align*}
		% \mathbf{H}=\left[
		% \begin{smallmatrix}
		% 	1 &-1 & 0 & 0\\
		% 	0 & 1 & 1 & 0\\
		% 	0 & 0 & 1 & -1\\
		% 	1 & 0 & 0 & 1
		% \end{smallmatrix}
		% \right].
  %   \end{align*}
    Here, $(a)$ holds because of the zeros in the matrix $\mathbf{H}$. When we have at least five groups of workers (i.e., $c\geq 5$), we can prove that this scheme has recovery threshold $= c+13$.
    	\begin{lemma}
		The recovery threshold of this scheme is $c+13$.
	\end{lemma}
	\begin{IEEEproof}
Our overall idea is to show that if the central node can receive at least five evaluations from at least five distinct groups, it can decode the desired result. Towards this end, suppose that the central node receives results from two workers in the group $\mathcal{G}_i$. From Definition \ref{def:grad}, there exists $\mathbf{g}^T=[g_0, \cdots, g_3]$ such that it has non-zero entries only corresponding to the two workers that return their results with the property that  $\mathbf{g}^T \mathbf{\tilde{h}}_{\tilde{p}} =1$ for columns $\mathbf{\tilde{h}}_{\tilde{p}}$ of $\bfH$.
%and satisfies \eqref{eqn:span}. \kyungrak{need to explicitly state $\mathbf{g}^T\mathbf{h}_i=1$ because I use this in calculating (4)} 
Then, we note that
\begin{align} 
&\sum_{w=0}^{3}g_{w}\mathbf{C}_w(x_i) \NNL
&= \sum_{w=0}^{3}\sum_{\tilde{p}=0}^{3}g_{w}h_{w,\tilde{p}}\bar{\A}^T(x_i,\tilde{p})\bar{\B}(x_i,\tilde{p})
    \NNL
   &= \sum_{\tilde{p}=0}^{3}\mathbf{g}^T \mathbf{\tilde{h}}_{\tilde{p}}\bar{\A}^T(x_i,\tilde{p})\bar{\B}(x_i,\tilde{p})\NNL
    &= %\stackrel{(a)}{=}
\sum_{\tilde{p}=0}^{3} \bar{\A}^T(x_i,\tilde{p})\bar{\B}(x_i,\tilde{p}) \label{eq:dec_w_g} \\
&= \sum_{\tilde{p}=0}^{3} \sum_{l_1 = 0}^2 \sum_{l_2=0}^2 x_i^{l_1 - l_2 + 2} \A^T_{3\tilde{p}+l_1}\B_{3\tilde{p} + l_2} \NNL
&= x_i^2 \sum_{l=0}^{11} \A^T_{l}\B_{l} + \text{interference terms} \label{eq:dec_final_exp}
%&=   
%\sum_{\tilde{p}=0}^{3} \left[x_i\left[\sum_{l=0}^{1} \A^T_{2\tilde{p}+l}\B_{2\tilde{p}+l} \right] +\A^T_{2\tilde{p}}\B_{2\tilde{p}+1}+x_i^2 \A^T_{2\tilde{p}+1}\B_{2\tilde{p}} \right] \NNL 
%& =  x_i^2 \!\!\underbrace{\left[\sum_{l=0}^{7}\A_{l}^T\B_{l}\right]}_{\textrm{useful term}} \!+\!  \underbrace{\left[\sum_{\tilde{p}=0}^{3} \A^T_{2\tilde{p}}\B_{2\tilde{p}+1}\right]\!+\! x_i^2 \left[\sum_{\tilde{p}=0}^{3}\A^T_{2\tilde{p}+1}\B_{2\tilde{p}}\right]}_{\textrm{interference term}}.	\label{eqn:exp_grad_multiplication}
		\end{align}
%where $\mathbf{\tilde{h}}_{\tilde{p}}$ is $\tilde{p}$-th column of the matrix $\mathbf{H}$. % and $I(x_i)$ is a polynomial of degree two for the variable $x_i$. 
    %Here, $(a)$ comes from \eqref{eqn:span}. \aditya{don't need this since we say it anyway before now.}
    %Since each row of $\mathbf{G}$ has at least a single zero, 
    Thus, we are able to obtain the useful term $\sum_{l=0}^{11}\A_{l}^T\B_{l}$ as the coefficient of $x_i^2$ in the above polynomial. Furthermore, note that the interference term does not depend on which workers returned their results since we are able to obtain \eqref{eq:dec_w_g} in the decoding process. %with arbitrary three different elements in each group. 
    Since the equation \eqref{eq:dec_final_exp} is a polynomial of degree four, we need at least five different interpolation points $x_i$ to obtain the useful term from the equation \eqref{eq:dec_final_exp}. Thus, obtaining five evaluations from five groups suffices to decode.
    
    To see that the recovery threshold is $c+13$, we proceed by contradiction. Note that, there are $c$ groups, each of which contains four nodes. It follows that we cannot decode when there are at most four groups where all the nodes return their results and all other groups are such that at most one node returns its result. Thus, we can have at most $4\times 4 + (c-4) = c+12$ nodes return their results in this case, i.e., decoding is guaranteed when $c+13$ workers return their results. 
    %Thus, in the worst case, we cannot decode when we receive only two different elements of $\mathcal{P}_{\mathsf{p}}$ from all groups, and then receive the remaining element of $\mathcal{P}_{\mathsf{p}}$ for just two different $x_i$. Therefore, we need at least $2\times c+2\times 2 +1=2c+5$ worker nodes.
	\end{IEEEproof} 
\end{example}

\subsection{General $k_A,k_B$ and $k_p$ \label{sec:general_scheme}}
We now consider the general case where each worker can store the equivalent of $\gamma_A=1/k_A k_p$ and $\gamma_B=1/k_B k_p$ fractions of matrices $\A$ and $\B$, respectively. %$\mu$ is an arbitrary positive integer such that $\mu|k_p$. 
In this case, the work of \cite{yu2020straggler} considers $k_p \times k_A$ and  $k_p  \times k_B$ block-decompositions of $\A$ and $\B$ respectively and proposes the EPC scheme with recovery threshold $k_p k_A k_B + k_p -1$. The encoding weight of the $\A$ and $\B$ matrices is $k_A k_p$ and $k_B k_p$ respectively.

Once again, in this case we are interested in schemes where the encoding weights of the $\A$ and $\B$ is lower and the degree of the polynomial that needs to be interpolated during decoding is lower.

For our scheme, we consider the following scenario. Let $m$ and $n$ be positive integers such that $k_A | m$ and $k_B | n$. Our scheme has another parameter $\Delta_p \leq k_p$ that allows us to tune the weight of the encoding. We set $p = LCM(\Delta_p, k_p)$. %p = \Delta_p k_p. 
As we saw in the motivating example, if $\Delta_p | k_p$, we will see that a simple scheme that essentially divides the overall scheme into $\frac{k_p}{\Delta_p}$ EP codes applies. Thus, for the discussion below it is instructive to consider the scenario where $\Delta_p$ does not divide $k_p$.

The central node first partitions the matrices $\A^T$ and $\B$ into submatrices as shown in \eqref{eq:block_decomp}.
% \begin{align*}
% \A^T
% %\!\!=\!\!
% &=
% \left[
% \begin{matrix}
% 	\A_{0,0}^T & \!\!\!\cdots \!\!\!& \A_{p-1,0}^T  \\
% 	\vdots & \!\!\!\ddots\!\!\! & \vdots  \\
% 	\A_{0,m-1}^T &\!\!\!\cdots \!\!\!& \A_{p-1,m-1}^T 
% \end{matrix}
% \right], \text{and}\\
% %
% %
% \B
% %\!=\!\!
% &=
% \left[
% \begin{matrix}
% 	\B_{0,0}  & \!\!\!\cdots\!\!\!& \B_{0,n-1} \\
%          \vdots & \!\!\!\ddots\!\!\! & \vdots  \\
% 	\B_{p-1,0}  &\!\!\!\cdots\!\!\! & \B_{p-1,n-1}
% \end{matrix}
% \right].
% \end{align*}
%Here $m,n$ and $p$, are chosen such that $k_A|m, k_B|n$, and $k_p|p$. Our scheme has another positive integer parameter $\Delta_p$ which is such that $\Delta_p | k_p$. Using this matrix block decomposition we present our scheme with storage fractions $\gamma_A=\mu/k_A k_p$ and $\gamma_B=\mu/k_B k_p$.  
We assume that there are $N= \frac{p}{\Delta_p} \cdot c$ workers, i.e., there are $c$ groups consisting of $\frac{p}{\Delta_p}$ nodes each. 

%where $\tilde{p} \in \mathcal{P}_{\mathsf{p}}=\{0, \cdots p-1\},\tilde{m} \in \mathcal{P}_{\mathsf{m}}$ and $\tilde{n} \in \mathcal{P}_{\mathsf{n}}$. 

%Since the submatrices of $\A^T$ and $\B$ has the portion of $1/pm$ and $1/pn$ of $\A$ and $\B$, respectively, 
The storage constraints imply that we can store the equivalent of $\frac{pm}{k_pk_A}$ encoded submatrices for $\A^T$ and $\frac{pn}{k_pk_B}$ encoded submatrices for $\B$ in each worker. Thus, we consider a worker group of $\frac{p}{\Delta_p}$ workers %$\mathcal{G}_i=\{W_{\frac{p}{\Delta_p}i}, \cdots, W_{\frac{p}{\Delta_p}(i+1)-1}\}$
$\mathcal{G}_i=\{\frac{p}{\Delta_p}i, \cdots, \frac{p}{\Delta_p}(i+1)-1\}$ for $i = 0, \dots, c-1$, where the $\frac{p}{\Delta_p}i+w$-th worker stores 
\begin{align*}
\bar{\A}^T(x_i,\tilde{p},\tilde{m}) &=
\sum_{l=0}^{\Delta_p-1}\sum_{s=0}^{k_A-1}x_i^{l+s\Delta_p}\A_{\Delta_p\tilde{p}+l,k_A\tilde{m}+s}^T,
\end{align*}
for all $\tilde{p}\in \{w,\cdots,w+ \frac{p}{k_p}-1\} \mod \frac{p}{\Delta_p}$ and $\tilde{m} \in \mathcal{P}_\mathsf{m}=\{0,\cdots,\frac{m}{k_A}-1\}$, 
and 
\begin{align*}
\bar{\B}(x_i,\tilde{p},\tilde{n}) &= 
\sum_{l=0}^{\Delta_p-1}\sum_{u=0}^{k_B-1}x_i^{\Delta_p-1-l+u\Delta_pk_A}\B_{\Delta_p\tilde{p}+l,k_B\tilde{n}+u},
\end{align*}
for all $\tilde{p}\in \{w,\cdots,w+\frac{p}{k_p}-1\} \mod \frac{p}{\Delta_p}$ and $\tilde{n} \in \mathcal{P}_\mathsf{n}=\{0,\cdots,\frac{n}{k_B}-1\}$. 
%Here, $\Delta_p$ is an integer which divides $k_p$ (i.e., $\Delta_p|k_p$). 
%
%Note that $\Delta_p$ controls the degree of the polynomial when $\bar{\A}^T$ and $\bar{\B}$ are multiplied\blue{, to control the numerical stability.}
Also, $x_i$ is the same for all workers in the group.

Now, we choose a gradient coding matrix ({\it cf.} Definition \ref{def:grad}) $\mathbf{H}$ with parameters $\eta=\frac{p}{\Delta_p}$ and $\kappa=\frac{p}{k_p}-1$.
Then, the $w$-th worker in the $i$-th group computes 
	\begin{align}
		\mathbf{C}_w (x_i,\tilde{m},\tilde{n}) \!&=\!
		\sum_{\tilde{p}=w}^{w+\frac{p}{k_p}-1}h_{w,\tilde{p}}\bar{\A}^T(x_i,\tilde{p},\tilde{m})\bar{\B}(x_i,\tilde{p},\tilde{n})\NNL
        &=\sum_{\tilde{p}=0}^{\frac{p}{\Delta_p}-1}h_{w,\tilde{p}}\bar{\A}^T(x_i,\tilde{p},\tilde{m})\bar{\B}(x_i,\tilde{p},\tilde{n}) \NN
	\end{align}
 for all $\tilde{m} \in \mathcal{P}_\mathsf{m}$ and $ \tilde{n}\in \mathcal{P}_\mathsf{n}$. The summation indices are reduced modulo $\frac{p}{\Delta_p}$ in the expression above. The last step above holds because of the properties of the gradient coding matrix.% Here, $h_{i,j}$ is the $i$th row and $j$th column of the matrix

\begin{table*}[t!]
\small
     \centering
    \begin{tabular}{|c|c|c|}
    \hline
        & Entangled Polynomial code 
        & Proposed 
        \\
    \hline\hline
       Recovery threshold & $k_A k_B k_p+k_p-1$ & { $\left(\dfrac{p}{\Delta_p}-\dfrac{p}{k_p}\right)\cdot c +\dfrac{p}{k_p} \cdot (k_Ak_B\Delta_p+\Delta_p-2) \!+\!1$} \\
    \hline
       Number of assignments per worker & 1 & $\frac{pmn}{k_pk_Ak_B}$ \\       
    \hline
       Computational cost per worker & $O\left(\frac{\alpha\gamma}{k_Ak_B} \cdot 2\frac{\beta}{k_p}\right)=O\left(\frac{2\alpha\beta\gamma}{k_pk_Ak_B}\right)$  & $O\left(\frac{\alpha\gamma}{mn} \cdot 2\frac{\beta}{p} \cdot \frac{pmn}{k_pk_Ak_B}\right) =O\left(\frac{2\alpha\beta\gamma}{k_pk_Ak_B}\right) $\\ 
    \hline 
     \rule{0pt}{3ex}    Encoding weight of $\bar{\A}$ (or $\bar{\B}$)& $k_pk_A$ (or $k_pk_B$) & $\Delta_pk_A$ (or $\Delta_pk_B$) \\
    \hline
    \end{tabular}
    \vspace{0.1in} \caption{Performance comparison of the various schemes}
    \vspace{-0.3in}
    \label{tab:exp3}
\end{table*}
% Finally, we can apply the entangled polynomial code to $\tilde{\mathbf{C}}_i, i \in \mathcal{I}$. Pick $k_Ak_B\Delta_p+\Delta_p-1$ indices from $\mathcal{I}$ and define it as $\hat{\mathcal{I}}$. Also, let $\mathbf{P}$ the Vandermode matrix with the parameters given as $\{x_i|i \in \hat{\mathcal{I}}\}$. 
% Then, for given $(\tilde{m},\tilde{n})\in \{0,1\cdots,\frac{p}{k_A}-1\} \times \{0,1\cdots,\frac{p}{k_B}-1\}$ and $(s,u)\in \{0,1\cdots,k_A-1\} \times \{0,1\cdots,k_B-1\}$, we can finally estimate the $(k_A\tilde{m}+s,k_B\tilde{n}+u)$-th desired term of $\A^T\B$ as 
% \begin{align*}
%     \hat{\mathbf{C}}_{k_A\tilde{m}+s,k_B\tilde{n}+u}
%     &=\sum_{l'=0}^{p-1} \A_{l',k_A\tilde{m}+s}^T\B_{l',k_B\tilde{n}+u}\\
%     &=\sum_{i \in \hat{\mathcal{I}}}(\mathbf{P}^{-1})_{\Delta_p-1+s\Delta_p+u\Delta_pk_A,i} \cdot \tilde{\mathbf{C}}_i
% \end{align*}
% using the entangled polynomial code \cite{Yu2020}. 

%Note that since $\mathbf{g}$ has zeros in $p/k_p$ positions, we can obtain the expression \eqref{eqn:grad_multipication} using results from $\frac{p}{\Delta_p}-\frac{p}{k_p}+1$ workers within the group.
Define $\tau_{\mathsf{GC-EPC}}$ as the recovery threshold of the proposed scheme. The subscript GC-EPC refers to the fact that we combine gradient coding and entangled polynomial coding in this approach. The proof of the following theorem appears in the Appendix.
\begin{theorem}\label{thm:Prop_RecThr}
For a given parameter $\Delta_p \leq k_p$, we need at least $c \geq k_Ak_B\Delta_p+\Delta_p-1$ worker groups and the recovery threshold of the scheme is 
\begin{equation}
\label{eqn:tau_grad}%=2c+5 
\begin{aligned}
    \tau_{\mathsf{GC-EPC}}&\!={ \left(\dfrac{p}{\Delta_p}-\dfrac{p}{k_p}\right)\cdot c +\dfrac{p}{k_p} \cdot (k_Ak_B\Delta_p+\Delta_p-2) \!+\!1}. %\NNL
    %&= (p-1)\cdot c + k_Ak_B\Delta_p+\Delta_p-1. 
\end{aligned}    
\end{equation}
\end{theorem}

%\blue{The precise decoding algorithm is summarized in Algorithm \ref{alg:decoding_alg}.}

 \begin{remark} 
 From the encoding scheme, we can clearly see that overall polynomial to be interpolated is now of degree $k_A k_B \Delta_p + \Delta_p - 2$ as opposed to $k_A k_B k_p + k_p -2$ for the EP code. Thus numerical stability improves. Next, the weight of the encoding of $\A$ and $\B$ matrices is $k_A \Delta_p$ and $k_B \Delta_p$ as against $k_A k_p$ and $k_A k_p$ for the EP code, respectively. Of course, these benefits trade-off with a worse recovery threshold.
 \end{remark}

\begin{example}
Consider a scenario where $k_A = k_B = 1$ and $k_p=15$. In this case, the EPC code has a threshold of $29$ and the encoding weight of both the $\A$ and $\B$ matrices is $15$. We note that interpolating a polynomial of degree $28$ will already result in significant numerical issues whereby the decoded result will essentially be useless (see Section \ref{sec:sim_results}). %\aditya{Should give numbers of MATLAB simulation here.}

For our scheme, suppose that we have $N \geq 5c$ workers and that we set $\Delta_p = 6$. Then, we will choose $p = LCM(6,15) = 30$. The corresponding threshold will be $3c + 21$, and the encoding weights for both the $\A$ and $\B$ matrices will be $6$. We note here that the decoder will only interpolate a polynomial of degree $10$ which is much smaller than the Mat-Dot code.
\end{example}
\subsection{Discussions \label{sub:discussion}}
%%%%%%%%%%%%%%%%%%%%%%%%%%%%%%%%%%%%%%%%%%%%%%%%%%%%%%%%%%%%%%%%%%%
\begin{table}[h] 
    \centering
    \begin{tabular}{c|cccc|cccc}
        \specialrule{.2em}{.1em}{.1em} 
         $N$ & $k_A,k_B \atop m,n$ & $k_p$& $\Delta_p$ & $p$ & $\tau_{\mathsf{epc}}$ & $\!\!\tau_{\mathsf{GC-EPC}}$ & $\!\!{\text{wt}}_{\mathsf{epc}}$ & $\!\!{\text{wt}}_{\mathsf{GC-EPC}}$ \\
         \hline
         24 & 1 & 6 & 4 & 12 & 11 & 21 & 6 & 4 \\%c+13 24/3
         24 & 1 & 6 & 3 & 6 & 11 & 17 & 6 & 3 \\%c+5 24/2
         24 & 1 & 6 & 2 & 6 & 11 & 19 & 6 & 2 \\%2c+3 24/3
         10 & 1 & 6 & 3 & 6 & \textrm{N/A} & 10 & 6& 3 \\ %c+5 10/2
         64 & 1 & 4 & 3 & 12 & 7 & 29 & 4 & 3 \\ %c+13 64/4
         64 & 1 & 8 & 3 & 24 & 15 & 53 & 8 & 3 \\ %5c+13 64/8
         64 & 2 & 4 & 3 & 12 & 19 & 56 & 8 & 6 \\ %c+40  64/4
        \specialrule{.2em}{.1em}{.1em} 
    \end{tabular}
    \caption{Simple comparison of the recovery threshold and number of weights with the various system parameters}
    \label{tab:comparison} \vspace{-0.3in}
\end{table}
%%%%%%%%%%%%%%%%%%%%%%%%%%%%%%%%%%%%%%%%%%%%%%%%%%%%%%%%%%%%%%%%%%%
A comparison of the various performance measures of the proposed scheme and the EPC scheme is summarized in Table \ref{tab:exp3}. % Furthermore, to clearly observe the performance of the proposed scheme, we add an example with specific parameters.  
%\begin{example}
The recovery threshold and the number of weights of the proposed scheme and the EPC scheme with various system parameters are discussed in Table \ref{tab:comparison}. In Table \ref{tab:comparison}, ${\text{wt}}_{\mathsf{epc}}, {\text{wt}}_{\mathsf{GC-EPC}}$, and $\tau_{\mathsf{epc}}, \tau_{\mathsf{GC-EPC}}$ denotes the encoding weights for EPC scheme and the proposed scheme, and the recovery threshold of the EPC scheme and the proposed scheme respectively. From Table \ref{tab:comparison}, we can observe that the threshold of the GC-EPC scheme is higher than the EPC scheme when $\Delta_p < k_p$. However, the encoding weights are lower. Moreover, as discussed shortly in Section \ref{sec:sim_results}, our scheme is much more numerically stable.

We observe that our scheme reduces to other well-known schemes for specific parameter regimes.
\begin{itemize}
    \item If $\Delta_p=p=k_p$, the proposed scheme is the same as the EPC scheme \cite{yu2020straggler} and the recovery threshold becomes $k_pk_Ak_B+k_p-1$.
    \item If $\Delta_p=1$, $m=n=1$ and $p=k_p$, the proposed scheme equals to the scheme which just applies GC to uncoded matrices (\textsf{CMM-1} scheme in \cite{Charalambides2021}) and the recovery threshold becomes $(p-1)c+1$.
    \item 	If $\Delta_p=1$ and $k_p=p=1$, the proposed scheme equals to the EPC scheme \cite{yu2020straggler} for $p=1$ (or \textsf{CMM-3} scheme in \cite{Charalambides2021}). The recovery threshold becomes $k_Ak_B$.
\end{itemize}

% Furthermore, considering three particular sparsity conditions and partitioning parameters (i.e, $\Delta_p,p,m$ and $n$), \cite{Yu2020} and \cite{Charalambides2021} can be viewed as extreme cases of the proposed scheme. 
% \begin{corollary}[when $\Delta_p=p=k_p$]
% 	If $\Delta_p=p=k_p$, the proposed scheme is the same as the EPC scheme \cite{Yu2020} and the recovery threshold becomes $k_pk_Ak_B+k_p-1$.
% \end{corollary}
% \begin{corollary}[when $\Delta_p=1$, $m=n=1$, and $p=k_p$]
% 	If $\Delta_p=1$, $m=n=1$ and $p=k_p$, the proposed scheme equals to the scheme which just applies GC to uncoded matrices (\textsf{CMM-1} scheme in \cite{Charalambides2021}) and the recovery threshold becomes $(p-1)c+1$.
% \end{corollary}
% \begin{corollary}[when $\Delta_p=1$ and $k_p=p=1$]
% 	If $\Delta_p=1$ and $k_p=p=1$, the proposed scheme equals to the EPC scheme \cite{Yu2020} for $p=1$ (or \textsf{CMM-3} scheme in \cite{Charalambides2021}). The recovery threshold becomes $k_Ak_B$.
% \end{corollary}
Note that all the three schemes \cite{yu2020straggler}, \cite{Charalambides2021}, and our proposed GC-EPC scheme have the same computational cost per worker.

\section{Simulation result}
\label{sec:sim_results}
In this section, we evaluate and compare our proposed schemes with benchmark schemes in terms of two different performance measures. 
\begin{itemize}
    \item First, we measure numerical stability caused by distributing the computations. i.e., the computation error of the reconstructed solution $\hat{\mathbf{C}}$ normalized by the original solution given by
\begin{align*}
    \dfrac{\Vert \hat{\mathbf{C}} - \mathbf{A}^T\mathbf{B}\Vert_F}{\Vert\mathbf{A}^T\mathbf{B}\Vert_F}.
\end{align*}
    \item Second, we compare the average computation time (in seconds), the computation time consumed until receiving the computation results from $\tau_{\star}, \star=\mathsf{epc}$ or $\mathsf{GC-EPC}$ number of worker nodes, of each scheme. 
\end{itemize}
For a fair comparison, we consider two different schemes for distributed matrix multiplication under the same storage capacity $\gamma_A=1/k_Ak_p$ for the matrix $\A^T$ and $\gamma_B=1/k_Bk_p$ for the matrix $\B$ as follows.
\begin{itemize}
    \item EPC scheme: The EPC scheme in \cite{yu2020straggler}, where the interpolation points are spaced equidistant on the interval $[-1,1]$. The matrices $\A^T$ and $\B$ is partitioned into $k_A \times k_p$ submatrices and $k_p \times k_B$ submatrices, respectively.
    %\item Conventional scheme 2: Gradient coding \cite{Tandon2017} applied to uncoded matrices (\textsf{CMM-1} scheme in \cite{Charalambides2021}). To satisfy the storage constraint, both the matrices $\A$ and $\B$ are partitioned into $k_p \times 1$ submatrices.
    \item Proposed scheme: The proposed scheme for general $k_A,k_B$, and $k_p$ is illustrated in subsection \ref{sec:general_scheme}. Similar to the EPC scheme, the interpolation points are randomly generated with real values having equal distance between $[-1,1]$. 
\end{itemize}

For the simulation environment, we consider the input matrices $\A$ and $\B$ having the size $5040 \times 5040$. We consider the storage parameter as $k_A=1,k_B=1$, where each matrix has the $0.01$ fraction of nonzero elements (i.e., sparsity parameter $\rho=0.01$), and $k_p=14$. The total number of workers is given as $N=420$.

%%%%%%%%%%%%%%%%%%%%%%%%%%%%%%%%%%%%%
\begin{figure}[t!]
		\centering		{\includegraphics[width=.43\textwidth]{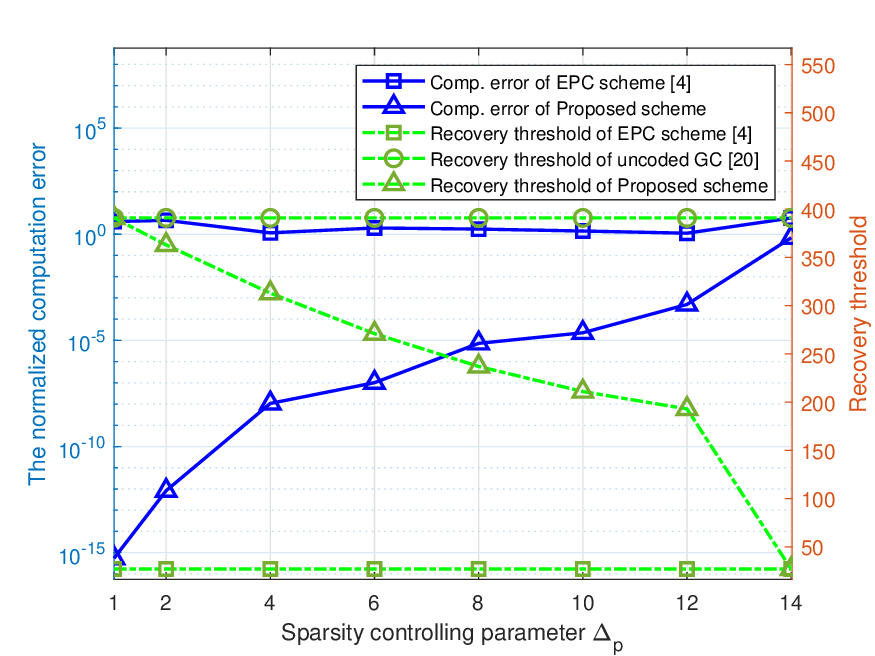} }
	\vspace{0.1in}
	\caption{A plot of the trade-off between computation error and recovery threshold of the various schemes with respect to $\Delta_p$.}
	\label{fig:tradeoff_graph}
\end{figure}
%%%%%%%%%%%%%%%%%%%%%%%%%%%%%%%%%%
In Fig. \ref{fig:tradeoff_graph}, we compare the normalized computation errors and recovery thresholds of various schemes with respect to the sparsity controlling parameter $\Delta_p$. The blue bold lines and the green dashed line represent the computation errors and recovery thresholds, respectively.
As we expected, we can first observe that the EPC scheme is numerically unstable. On the other hand, the proposed scheme can provide numerical stability while still having straggler resilience. 
Also, we can observe the tradeoff between the computation error and the recovery threshold. i.e.,
the computation error of the proposed scheme increases and the recovery threshold decreases as the sparsity controlling parameter $\Delta_p$ increases. 
Finally, we can observe that the recovery threshold of the proposed scheme meets the EPC scheme and GC scheme in extreme cases as we discussed in subsection \ref{sub:discussion}.

%%%%%%%%%%%%%%%%%%%%%%%%%%%%%%%%%%%%%
\begin{figure}[t!]
		\centering		{\includegraphics[width=.43\textwidth]{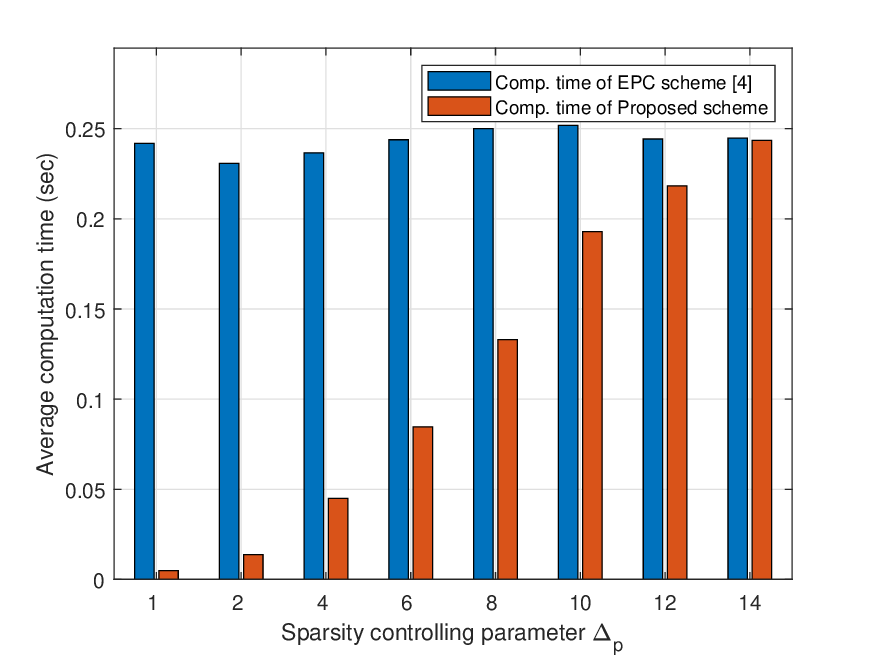} }
	\vspace{0.1in}
	\caption{The average computation time of various schemes with respect to $\Delta_p$.}
	\label{fig:comp_time}
\end{figure}
%%%%%%%%%%%%%%%%%%%%%%%%%%%%%%%%%%
In Fig. \ref{fig:comp_time}, we compare the average computation time of the EPC scheme and the proposed scheme with respect to sparsity controlling parameter $\Delta_p$. We can observe that the proposed scheme is faster than the EPC scheme. This is because the number of summation terms of $\bar{\A}$ and $\bar{\B}$ is reduced in the proposed scheme, so the sparsity of the encoded matrices for the proposed scheme is preserved compared to that for the EPC scheme. Since sparse matrix multiplication is faster than dense matrix multiplication, we can observe that the proposed scheme is faster than the EPC scheme.

\appendices

\section{Proof of Theorem \ref{thm:Prop_RecThr}}
\begin{IEEEproof}
We show that if the central node can receive at least $\frac{p}{\Delta_p}-\frac{p}{k_p}+1$ evaluations from at least $k_Ak_B\Delta_p+\Delta_p-1$ groups, it can decode the desired result. Towards this end, suppose that the central node receives results from $\frac{p}{\Delta_p}-\frac{p}{k_p}+1$ workers in the group $\mathcal{G}_i$. 

From Definition \ref{def:grad}, there exists $\mathbf{g}^T=[g_{0}, \cdots, g_{\frac{p}{\Delta_p}-1}]$ such that it has non-zero entries corresponding to the index of $\frac{p}{\Delta_p}-\frac{p}{k_p}+1$ workers that return their results with the property that  $\mathbf{g}^T \mathbf{\tilde{h}}_{\tilde{p}} =1$ for columns $\mathbf{\tilde{h}}_{\tilde{p}}$ of $\bfH$.  Then, we note that we can obtain the useful term %added with some interference terms 
 from the following linear combination of received computation results.
 \begin{align}
     \tilde{\mathbf{C}}_i(\tilde{m},\tilde{n})
      % &=\sum_{w \in \hat{\mathcal{G}}_i} g_{w}\mathbf{C}_w(x_i,\tilde{p},\tilde{m},\tilde{n})\\&\stackrel{(a)}{=}
    &=\sum_{w=0}^{\frac{p}{\Delta_p}-1}g_{w}\mathbf{C}_w(x_i,\tilde{m},\tilde{n})\NNL
    &= 
    \sum_{w=0}^{\frac{p}{\Delta_p}-1}\sum_{\tilde{p}=0}^{\frac{p}{\Delta_p}-1}g_{w}h_{w,\tilde{p}}\bar{\A}^T(x_i,\tilde{p},\tilde{m})\bar{\B}(x_i,\tilde{p},\tilde{n})\NNL
    &=\sum_{\tilde{p}=0}^{\frac{p}{\Delta_p}-1}\mathbf{g}^T\mathbf{\tilde{h}}_{\tilde{p}}\bar{\A}^T(x_i,\tilde{p},\tilde{m})\bar{\B}(x_i,\tilde{p},\tilde{n}) \NNL
    &= \sum_{\tilde{p}=0}^{\frac{p}{\Delta_p}-1}\bar{\A}^T(x_i,\tilde{p},\tilde{m})\bar{\B}(x_i,\tilde{p},\tilde{n}), \label{eqn:no_depend_w}
 \end{align}
 which is equivalent to \eqref{eqn:grad_multipication} shown at the top of the page, for each $\tilde{m} \in \mathcal{P}_\mathsf{m}$ and $\tilde{n}\in \mathcal{P}_\mathsf{n}$. We note that for fixed $\tilde{m}$ and $\tilde{n}$ the right hand side (RHS) of \eqref{eqn:grad_multipication} contains $k_A k_B$ useful terms. 

 \begin{figure*}
    \begin{equation}
\label{eqn:grad_multipication}
    \begin{aligned}	\sum_{\tilde{p}=0}^{\frac{p}{\Delta_p}-1}\!\!\bar{\A}^T(x_i,\tilde{p},\tilde{m})\bar{\B}(x_i,\tilde{p},\tilde{n}) 
    &=  \sum_{\tilde{p}=0}^{\frac{p}{\Delta_p}-1}
    \left[\sum_{l_1=0}^{\Delta_p-1}\sum_{l_2=0}^{\Delta_p-1}\sum_{s=0}^{k_A-1}\sum_{u=0}^{k_B-1} \left[x_i^{\Delta_p-1+s\Delta_p+u\Delta_pk_A+l_1-l_2} \cdot \A_{\Delta_p\tilde{p}+l_1,k_A\tilde{m}+s}^T\B_{\Delta_p\tilde{p}+l_2,k_B\tilde{n}+u}\right]\right] \\    
    &=  \sum_{\tilde{p}=0}^{\frac{p}{\Delta_p}-1}
    \left[\sum_{s=0}^{k_A-1}\sum_{u=0}^{k_B-1} \left[x_i^{\Delta_p-1+s\Delta_p+u\Delta_pk_A} \cdot \sum_{l=0}^{\Delta_p-1}\A_{\Delta_p\tilde{p}+l,k_A\tilde{m}+s}^T\B_{\Delta_p\tilde{p}+l,k_B\tilde{n}+u}\right]\right]
    \!\!+ \mathsf{Int}(x_i) \\
    & = \sum_{s=0}^{k_A-1}\sum_{u=0}^{k_B-1} \left[ x_i^{\Delta_p-1+s\Delta_p+u\Delta_pk_A} \cdot \underbrace{\sum_{l'=0}^{p-1} \A_{l',k_A\tilde{m}+s}^T\B_{l',k_B\tilde{n}+u}}_{\textrm{useful term}} \right]+\mathsf{Int}(x_i)
\end{aligned}
\end{equation}
\hrule
\end{figure*}
 
 Furthermore, $\mathsf{Int}(x_i)$ is a remaining interference term which is a polynomial of degree $k_Ak_B\Delta_p+\Delta_p-2$ for the variable $x_i$. We provide the exact expression of $\mathsf{Int}(x_i)$ at Appendix \ref{App:Int_terms}.
 %\blue{Also, $(a)$ comes from \eqref{eqn:span}. %$\mathbf{g}^T\mathbf{H}=\mathds{1}^{1 \times \frac{p}{\Delta_p}}$.}
%\aditya{some grammar and rewording issues here}
Now, in order to guarantee that we can decode the desired terms from $\tilde{\mathbf{C}}_i(\tilde{m},\tilde{n})$, we need to verify the following conditions
\begin{itemize}
\item The form of $\tilde{\mathbf{C}}_i(\tilde{m},\tilde{n})$ is the same regardless of which (at least) $\frac{p}{\Delta_p}-\frac{p}{k_p}+1$ workers in a group return their results. This is equivalent to asserting that the terms have no dependence on the index $w$.
\item The desired terms and the interference terms appear as coefficients of different degree terms in $\tilde{\mathbf{C}}_i(\tilde{m},\tilde{n})$.
\item All desired terms appear as coefficients of different degrees in $\tilde{\mathbf{C}}_i(\tilde{m},\tilde{n})$.
\end{itemize}

%i) the interference terms has no dependence on $w$, ii) the useful terms and the interference terms lies in the coefficients of different degrees in $\tilde{\mathbf{C}}_i(\tilde{m},\tilde{n})$, and iii) all the useful terms lies in the coefficients of different degrees in $\tilde{\mathbf{C}}_i(\tilde{m},\tilde{n})$. 

We provide detailed proof of these claims in Appendix \ref{App:Prop_RecThr}.

%\aditya{General comment: The proof of correctness has three part - (i) the term has no dependence on w, so the interference term remains the same regardless of which workers return their results in a group, (ii) the interference terms and the useful terms have different degrees and (iii) all useful terms appear at different degrees. All these claims need to be made precisely.}

Since the equation \eqref{eqn:grad_multipication} is a polynomial of degree $k_Ak_B\Delta_p+\Delta_p-2$, we need at least $k_Ak_B\Delta_p+\Delta_p-1$ different interpolation points $x_i$ to extract the useful terms from the equation \eqref{eqn:grad_multipication}. Thus, in order to have $k_Ak_B\Delta_p+\Delta_p-1$ different $x_i$, we need at least $k_Ak_B\Delta_p+\Delta_p-1$ worker groups. Thus, to see that the recovery threshold is \eqref{eqn:tau_grad}, we proceed by contradiction. Note that, there are $c$ groups, each of which contains $\frac{p}{\Delta_p}$ nodes. Also, we need $\frac{p}{\Delta_p}-\frac{p}{k_p}+1$ for each worker group to obtain the equation \eqref{eqn:grad_multipication}. It follows that we cannot decode when there are at most $k_Ak_B\Delta_p+\Delta_p-2$ groups where all the nodes return their results and all other groups are such that at most $\frac{p}{\Delta_p}-\frac{p}{k_p}$ nodes return their results. 
% in the worst case, we cannot decode when we receive $\mathbf{C}_w (x_i,\tilde{p},\tilde{m},\tilde{n})$ for only $\frac{p}{\Delta_p}-\frac{p}{k_p}$ different elements of $w \in \mathcal{P}_{\mathsf{p}}=\{0,\cdots,\frac{p}{\Delta_p}-1\}$ from all groups, and then receive the remaining element of $\mathcal{P}_{\mathsf{p}}$ for just $k_Ak_B\Delta_p+\Delta_p-2$ different groups. 
Therefore, we need at least the number of workers specified in \eqref{eqn:tau_grad} to guarantee the decoding.
\end{IEEEproof}
\vspace{5mm}
\section{Interference term of the equation \eqref{eqn:grad_multipication} \label{App:Int_terms}}
\begin{align*}
    &\sum_{\tilde{p}=0}^{\frac{p}{\Delta_p}-1}
    \left[\sum_{l_1= 0}^{\Delta_p-1}\sum_{l_2= 0 \atop l_2 \neq l_1}^{\Delta_p-1}\sum_{s=0}^{k_A-1}\sum_{u=0}^{k_B-1} \left[x_i^{\Delta_p-1+s\Delta_p+u\Delta_pk_A+l_1-l_2}\right.\right.\\
    &\qquad \left.\rule{0em}{8mm} \left.\cdot \A_{\Delta_p\tilde{p}+l_1-l_2+l_2,k_A\tilde{m}+s}^T\B_{\Delta_p\tilde{p}+l_2,k_B\tilde{n}+u}\right]\right]\\
    &= \sum_{\tilde{l}= -\Delta_p+1 \atop \tilde{l} \neq 0}^{\Delta_p-1}\sum_{s=0}^{k_A-1}\sum_{u=0}^{k_B-1} \left[\sum_{\tilde{p}=0}^{\frac{p}{\Delta_p}-1}\sum_{l_2=0}^{\Delta_p-1} x_i^{\Delta_p-1+s\Delta_p+u\Delta_pk_A+\tilde{l}}\right.\\
    &\qquad \left.\rule{0em}{7mm} \cdot \A_{\Delta_p\tilde{p}+l_2+\tilde{l},k_A\tilde{m}+s}^T\B_{\Delta_p\tilde{p}+l_2,k_B\tilde{n}+u} \right]\\
    &=\sum_{\tilde{l}= -\Delta_p+1 \atop \tilde{l} \neq 0}^{\Delta_p-1}\sum_{s=0}^{k_A-1}\sum_{u=0}^{k_B-1} \left[x_i^{\Delta_p-1+s\Delta_p+u\Delta_pk_A+\tilde{l}}\rule{0em}{7mm} \right.\\
    & \qquad \left.\cdot \sum_{l'=0}^{\Delta_p-1} \A_{l'+\tilde{l},k_A\tilde{m}+s}^T\B_{l',k_B\tilde{n}+u} \right]
    %\sum_{\tilde{p}=0}^{\frac{p}{\Delta_p}-1}
    %\left[\sum_{l'=-\Delta_p+1, \atop l'\neq 0}^{\Delta_p-1}\sum_{s=0}^{k_A-1}\sum_{u=0}^{k_B-1} \left[x_i^{\Delta_p-1+s\Delta_p+u\Delta_pk_A+l'} \cdot \A_{\Delta_p\tilde{p}+l_1,k_A\tilde{m}+s}^T\B_{\Delta_p\tilde{p}+l_2,k_B\tilde{n}+u}\right]\right]
\end{align*}
\vspace{5mm}

\section{Proof of the claims in Theorem \ref{thm:Prop_RecThr} \label{App:Prop_RecThr}}
First, note that, from \eqref{eqn:grad_multipication}, we can make the following observations.
\begin{itemize}
    \item Useful terms $\sum_{l=0}^{p-1} \A_{l,k_A\tilde{m}+s}^T\B_{l,k_B\tilde{n}+u}$ lie in the coefficients of $x_i^{\Delta_p-1+s\Delta_p+u\Delta_pk_A}$ for $(s,u) \in \{0,1\cdots,k_A-1\} \times \{0,1\cdots,k_B-1\}$ and $(\tilde{m},\tilde{n})\in \mathcal{P}_m \times \mathcal{P}_n$.
    \item All the other interference terms lie in the coefficients of $x_i^{\Delta_p-1+s\Delta_p+u\Delta_pk_A+\tilde{l}}$ for $(s,u) \in \{0,1\cdots,k_A-1\} \times \{0,1\cdots,k_B-1\}$ and $\tilde{l} \in \{\pm 1, \cdots, \pm(\Delta_p-1)\}$
\end{itemize}
Now, we can easily show the first claim. Since the equation \eqref{eqn:no_depend_w} has no dependence on $w$, not only the useful terms, but the interference terms also remain the same regardless of which workers return their results in a group. This is important in terms of the recovery threshold, otherwise, if the interference terms vary in $w$, we may still be able to decode the desired terms using $\tilde{\mathbf{C}}_i(\tilde{m},\tilde{n})$, but we need more workers than the degree of the polynomial to distinguish the interference terms. %(\red{I write down this part more because readers might not understand why making interference terms the same regardless of chosen workers is needed here.}) 
\begin{claim}
    The useful terms and the interference terms lie in the coefficients of different degrees in $\tilde{\mathbf{C}}_i(\tilde{m},\tilde{n})$.
\end{claim}
\begin{IEEEproof}
    We prove this by contradiction. Suppose there exist $(s_1,u_1), (s_2,u_2) \in \{0,\cdots, k_A-1\} \times \{0,\cdots, k_B-1\}$ and $\tilde{l} \in \{\pm 1, \cdots, \pm(\Delta_p-1)\}$ such that 
    \begin{align*}
        s_1\Delta_p+\Delta_p-1+u_1\Delta_pk_A = s_2\Delta_p+\Delta_p-1+u_2\Delta_pk_A+\tilde{l}.
    \end{align*}
    This further implies that
    \begin{align*}
        (s_1-s_2)\Delta_p+(u_1-u_2)\Delta_pk_A=\tilde{l}.
    \end{align*}
    Since the left hand side (LHS) is the multiple of $\Delta_p$ and the RHS is nonzero with an absolute value smaller than $\Delta_p$ we arrive at a contradiction. Thus, the claim holds.    
\end{IEEEproof}
\begin{claim}
    All the useful terms lie in the coefficients of different degrees in $\tilde{\mathbf{C}}_i(\tilde{m},\tilde{n})$.
\end{claim}
\begin{IEEEproof}
    We use contradiction again. Suppose there exist $(s_1,u_1), (s_2,u_2) \in \{0,\cdots, k_A-1\} \times \{0,\cdots, k_B-1\}$ such that 
    \begin{align*}
        s_1\Delta_p+\Delta_p-1+u_1\Delta_pk_A = s_2\Delta_p+\Delta_p-1+u_2\Delta_pk_A,
    \end{align*}
    $s_1\neq s_2$ and $u_1\neq u_2$.
    By some manipulations, this equation is equivalent to
    \begin{align*}
        (s_1-s_2)=(u_2-u_1)k_A.
    \end{align*}
    Since the RHS is the multiple of $k_A$ and the RHS is nonzero with an absolute value smaller than $k_A$. This is a contradiction. Thus, the claim holds.  
\end{IEEEproof}

\end{document}